# A Computational Approach to Finding RNA Tertiary Motifs in Genomic Sequences

Kevin Byron and Jason T. L. Wang[*]

Department of Computer Science, New Jersey Institute of Technology, University Heights, Newark, New Jersey 07102, USA

[*]Address correspondence to this author at the New Jersey Institute of Technology, Newark, New Jersey 07102, USA; Tel: 973-596-3396; E-mail: wangj@njit.edu

**ABSTRACT**

Motif finding in DNA, RNA and proteins plays an important role in life science research. Recent patents concerning motif finding in the biomolecular data are recorded in the DNA Patent Database which serves as a resource for policy makers and members of the general public interested in fields like genomics, genetics and biotechnology. In this paper we present a computational approach to mining for RNA tertiary motifs in genomic sequences. Specifically we describe a method, named CSminer, for finding RNA coaxial helical stackings in genomes. A coaxial helical stacking occurs in an RNA tertiary structure where two separate helical elements form a pseudocontiguous helix and provides thermodynamic stability to the molecule as a whole. Experimental results demonstrate the effectiveness of our approach.

**Keywords:** coaxial helical stacking, genome-wide motif finding, RNA junction.

**INTRODUCTION**

Motif finding in DNA, RNA and proteins plays an important role in life science research. Table 1 lists some recent patents concerning motif finding in the biomolecular data. These patents are recorded in the DNA Patent Database [1]. Here we present a computational approach to mining for RNA tertiary motifs in genomic sequences. Specifically we describe a method, named CSminer (i.e. Coaxial helical Stacking miner), for finding coaxial helical stackings in genomes. A coaxial helical stacking occurs in an RNA tertiary structure where two separate helical elements form a pseudocontiguous helix [2]. Coaxial helical stacking motifs occur in several large RNA structures, including tRNA [3], pseudoknots [4], group II intron [5] and large ribosomal subunits [6][7][8]. Coaxial helical stackings provide thermodynamic stability to the molecule as a whole [9][10], and reduce the separation between loop regions within junctions [11]. Moreover, coaxial helical stacking interactions form cooperatively with long-range interactions in many RNAs [12] and are thus essential features that distinguish different junction topologies.

Research to unravel the mysteries of (non-coding) RNA is exciting. An unexpected preliminary result of the human ENCODE project indicates that whereas protein-coding sequences (i.e. coding RNA) occupy less than 2% of the human genome, close to 93% of the genome is transcribed into non-coding RNA [13]. The "RNA World" hypothesis proposes that life based on RNA pre-dates the current world of life based on DNA, RNA and proteins [14]. Specialized RNA literature continually emerges [15]. The function of RNA is believed to be closely associated with its 3D structure, which, by virtue of canonical Watson-Crick base pairing (i.e. AU, GC) and wobble base pairing (i.e. GU), is largely determined by its 2D structure [16][17][18]. Many 2D structure prediction tools are available. One of the more highly regarded of these tools is Infernal [19] which has been, and continues to be, frequently cited [20][21][22][23]. Infernal applies stochastic context-free grammar methodology to efficiently predict 2D structures in genome-wide searches [24][25][26]. Databases detailing the 3D structure and features of RNA continue to grow [27][28]. Special interest is paid to RNA junctions [29][30] in which there are one or more coaxial helical stackings [31][32][33]. Statistical analysis approaches, in particular ensemble-based approaches, have been successful in non-life sciences applications [34][35]. Recently, these ensemble-based approaches have been successful in the field of bioinformatics [36][37][38][39][40][41][42]. Our Junction Explorer tool applies an ensemble-based approach, namely random forests, to predict the existence of a coaxial helical stacking in 3-way and higher-order RNA junctions [2]. In this work, we extend the functionality of Infernal to create a tool, named CSminer, which can efficiently predict the existence of coaxial helical stacks in genomes. This is accomplished by invoking Junction Explorer within Infernal and filtering Infernal results appropriately. Changes to the Infernal source code are available from the authors upon request.

| Patent No | Inventor | Title | Issue Date | U.S. Class | Ref # |
|---|---|---|---|---|---|
| 7884196 | Lawless | Vaccine composition comprising methylated DNA and immunomodulatory motifs | 2011-02-08 | 536/23.1 | [42] |
| 7622567 | Seeman, et al. | Multidimensional organization of heteromolecules by robust DNA motifs | 2009-11-24 | 536/23.1 | [43] |
| 7294494 | Roca | Nucleic acids encoding mutants of MAW motifs of RecA protein homologs | 2007-11-13 | 435/193 | [44] |
| 7279324 | Barak, et al. | Nucleic acid encoding G-protein coupled receptor with modified DRY motif | 2007-10-09 | 435/320.1 | [45] |
| 7101712 | Kondorosi, et al. | Plant protein with repeated WD40 motifs, nucleic acid coding for said protein, and uses thereof | 2006-09-05 | 435/468 | [46] |
| 6919438 | Alliel, et al. | Nucleic sequence and deduced protein sequence family with human endogenous retroviral motifs, and their uses | 2005-07-19 | 536/23.1 | [47] |
| 6774213 | Roca | Mutants of MAW motifs of RecA protein homologs, methods of making them, and their uses | 2004-08-10 | 530/350 | [48] |
| 6355426 | Prescott | Methods for the characterization and selection of RNA target motifs that bind compounds of pharmaceutical use | 2002-03-12 | 435/6 | [49] |
| 6300483 | Ludwig, et al. | Compositions inducing cleavage of RNA motifs | 2001-10-09 | 536/23.1 | [50] |

**Table 1.** Selected Patents for Motif Finding in Biomolecular Data

**MATERIAL AND METHODS**

**3-Way Junction Data Set**

For this work, we selected samples from known RNA 3-way junctions [19]. In [2], we present a table of attribute-value pairs describing 110 distinct RNA 3-way junctions confirmed in available crystal structures. Each 3-way junction contains a multi-branch loop (i.e. MBL) within the RNA molecule. An MBL is a naturally occurring structure in an RNA molecule and represents a junction of three or more helices. Each unique junction is assigned a sequential identifier, i.e. **Serial**, ranging from 0 through 109. The crystal structure containing the junction is identified by the attribute **PDB** representing an ID in PDB [27]. The type of RNA molecule is shown by the **RNA Type** attribute. The topology of each 3-way junction is identified by the **Family** code A, B or C [51]. The coaxial helical stacking status of each junction is given by the **Coaxial** attribute value which identifies the stacks, or helices, that share a common axis. A helix is represented in the form "Hn" where "n" is a number from 1 to 3 for each unique helix in the 3-way junction. H1 represents the helix whose base pairs contain the bases (i.e. nucleotides) closest to the 5' end and the 3' end of the RNA molecule. The H2 helix is the next helix encountered downstream in the 5' to 3' direction from the H1 helix. The H3 helix is the third helix in the junction.

The 3-way junction is fully described by attributes representing three RNA subsequences. For each subsequence, base coordinates and base values (i.e. A, C, G, U) are given. The starting and ending coordinates of the first subsequence are called **S1ID5** and **S1ID3** indicating the 5' and 3' ends of the first subsequence. Similarly, the coordinates of the second subsequence are called **S2ID5** and **S2ID3**, and the coordinates of the third subsequence are called **S3ID5** and **S3ID3**. The 3-way junction formed by these three subsequences includes unpaired bases of the MBL, terminal base pairs of the three helices and the penultimate (i.e. "next-to-last") base pairs of the three helices, as follows. The 5' end of the first subsequence is the 5' base of the penultimate base pair of (helix) H1. The 3' end of the first subsequence is the 5' base of the penultimate base pair of H2. Similarly, the 5' end of the second subsequence is the 3' base of the penultimate base pair of H2 and the 3' end of the second subsequence is the 5' base of the penultimate base pair of helix H3. It follows that the 5' end of the third subsequence is the 3' base of the penultimate base pair of H3 and the 3' end of the third subsequence is the 3' base of the penultimate base pair of helix H1.

The length of each subsequence is at least 4. The first two bases of a subsequence are part of one helix and the last two bases of that subsequence are part of the next sequential helix. There are zero or more unpaired bases between the two helices that share a subsequence. Unpaired bases of the first subsequence, as well as the number of these bases, are shown as attribute **J12**. Similarly, unpaired bases of the second subsequence, as well as the number of these bases, are shown as attribute **J23**, and unpaired bases of the third subsequence, as well as the number of these bases, are shown as attribute **J31**. Note that, generally, in helix Family A, the number of bases in **J31** is smaller than that in **J23**; in Family B, the number of bases in **J31** and **J23** is the same, and in Family C, the number of bases in **J31** is greater than that in **J23**. Finally, all bases comprising the first, second and third subsequences are called **StrSeq1**, **StrSeq2** and **StrSeq3**, respectively.

As an example, examine the 3-way junction identified as **Serial** 84, with **PDB** 1E8O whose **RNA Type** is ALU domain SRP (signal recognition particle). This 3-way junction has a coaxial helical stacking (**Coaxial**) identified as H1H3, i.e. helices H1 and H3 share a common axis. The RNA segment from position 100 (**S1ID5**) through 148 (**S3ID3**) is shown graphically in the figures below. Figure 1, obtained using RNAview [52], illustrates helices H1 and H3 aligned with a common axis. In addition to the canonical Watson-Crick base pairings (i.e. AU, GC) and wobble base pairings (i.e. GU), Figure 1 also illustrates tertiary interactions including pseudoknots. The primary sequence of RNA chain E obtained from PDB with highlighted 3-way junction subsequences is shown in Figure 2. The 2D structure plot for this RNA segment from position 100 through 148 is shown in Figure 3, obtained using S2S [53] and VARNA [54].

In Figure 3, a 3-way junction is enclosed within a red dotted line. The first subsequence comprising the 3-way junction starts at position 102 (5'), ends at position 105 (3') and consists of the bases CCGG. The second subsequence comprising the 3-way junction starts at position 122 (5'), ends at position 129 (3') and consists of the bases CUGUAGUC. Finally, the third subsequence comprising the 3-way junction starts at position 142 (5'), ends at position 145 (3') and consists of the bases GAGG. Unpaired bases in the MBL are those bases not part of the terminal base pairs of the 3 helices.

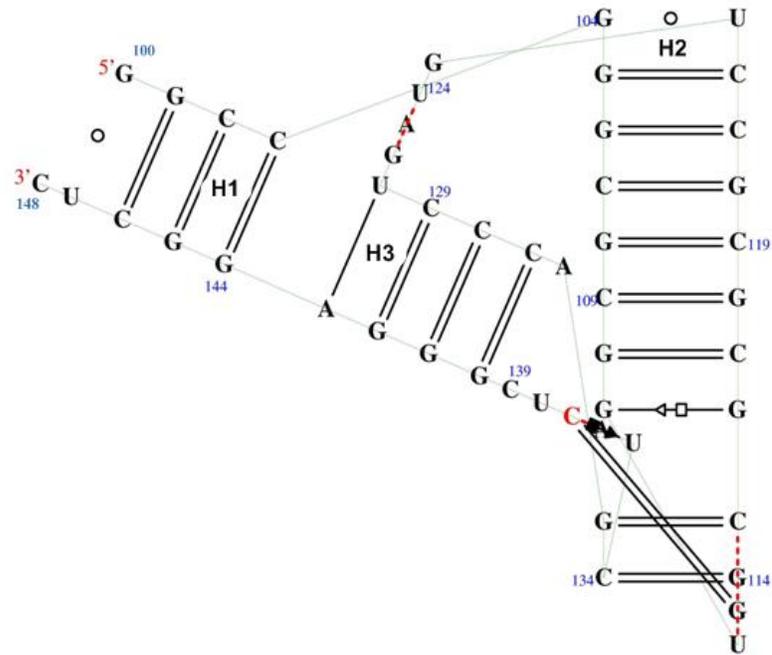

**Fig. (1)**. 2D (including tertiary interactions) illustration of bases 100 through 148 of RNA chain E from PDB ID 1E8O.

GG<mark>CCGG</mark>GCGCGGUGGCGCGCGC<mark>CUGUAGUC</mark>CCAGCUACUCGG<mark>GAGG</mark>CUC

**Fig. (2)**. Primary sequence of RNA chain E from PDB ID 1E8O illustrated in Figures 3 and 4. Highlighted in yellow are the three subsequences that comprise the 3-way junction.

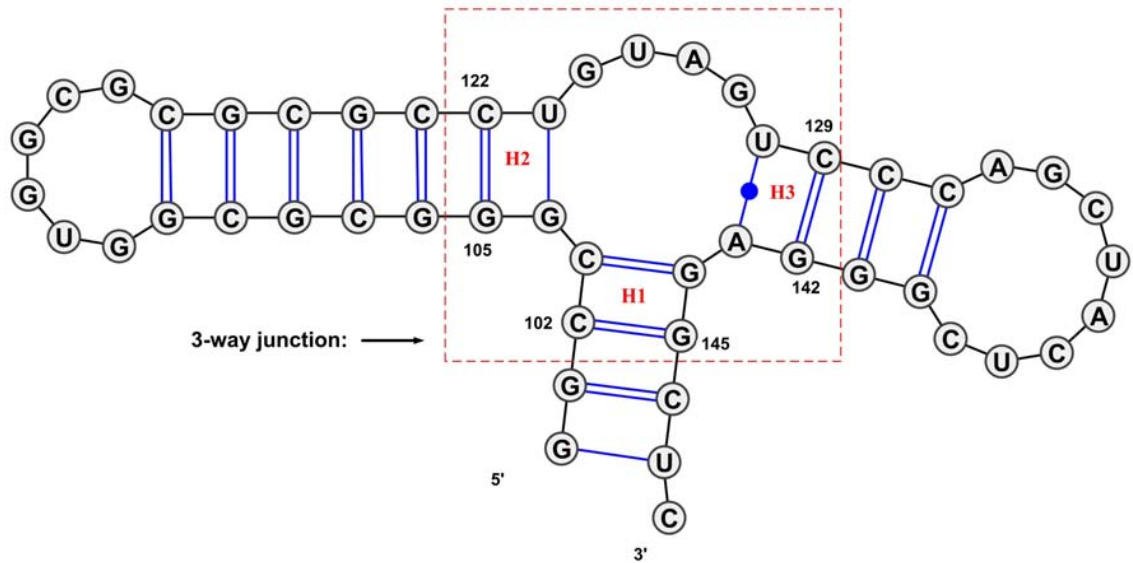

**Fig. (3)**. 2D plot produced by VARNA [54] using primary sequence of bases 100 through 148 of PDB ID 1E8O in CT format provided by S2S [53]. 3-way junction is enclosed by a dotted red line.

Figure 4, obtained using Jmol [55], presents a true 3D representation of the same RNA molecule. This representation is based on the crystal structure 3D coordinates of the 1,050 atoms comprising this RNA molecule. In this illustration, helix H1 is colored red, H2 is colored yellow and H3 is colored blue. The coaxial helical stacking of H1 and H3 is apparent in this illustration. The Jmol software allows the user to view a 3D visual rotation of the figure. By viewing the rotating figure, the helical stacking becomes even more apparent.

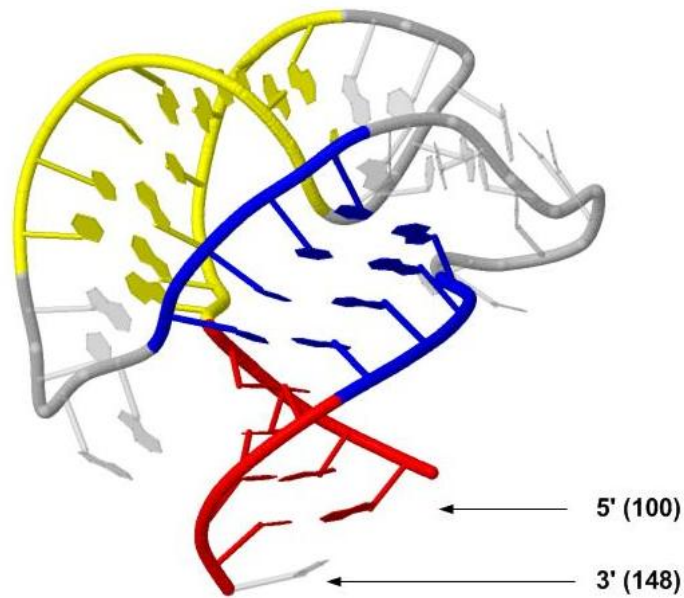

**Fig. (4)**. 3D plot produced by Jmol [55] using 3D coordinates of bases 100 through 148 of PDB ID 1E8O. Helix H1 is colored red, H2 is colored yellow and H3 is colored blue.

**3-Way Junction Feature Set**

A 3-way junction can be described using numerous and varied "features". Selecting an appropriate subset of these features, e.g. in motif prediction, is one of the most fundamental problems in bioinformatics, pattern recognition and machine learning. Table 2 identifies 15 features of 3-way junctions used to classify each junction according to its coaxial helical stacking status.

As an example, examine the 3-way junction located between bases 100 and 148 within chain E of the PDB 1E8O macromolecule. As previously mentioned, this 3-way junction has a coaxial helical stacking identified as H1H3, i.e. helices H1 and H3 share a common axis. Feature values for this 3-way junction are shown in Table 3 and can also be seen in the 2D plot illustrated in Figure 3.

| Feature | Description |
|---|---|
| $|J_{12}|$ | Number of nucleotides in the loop region between helix $H_1$ and helix $H_2$ |
| $|J_{23}|$ | Number of nucleotides in the loop region between helix $H_2$ and helix $H_3$ |
| $|J_{13}|$ | Number of nucleotides in the loop region between helix $H_1$ and helix $H_3$ |
| $Min(|J_{12}|,|J_{23}|,|J_{13}|)$ | The minimum value of $|J_{12}|$, $|J_{23}|$ and $|J_{13}|$ |
| $Med(|J_{12}|,|J_{23}|,|J_{13}|)$ | The median value of $|J_{12}|$, $|J_{23}|$ and $|J_{13}|$ |
| $Max(|J_{12}|,|J_{23}|,|J_{13}|)$ | The maximum value of $|J_{12}|$, $|J_{23}|$ and $|J_{13}|$ |
| $Min(|J_{23}|,|J_{13}|)$ | Minimum value of $|J_{23}|$ and $|J_{13}|$ |
| $Min(|J_{12}|,|J_{13}|)$ | Minimum value of $|J_{12}|$ and $|J_{13}|$ |
| $Min(|J_{12}|,|J_{23}|)$ | Minimum value of $|J_{12}|$ and $|J_{23}|$ |
| $A(J12)$ | Maximum number of consecutive adenines in the loop region between helix $H_1$ and helix $H_2$ |
| $A(J23)$ | Maximum number of consecutive adenines in the loop region between helix $H_2$ and helix $H_3$ |
| $A(J13)$ | Maximum number of consecutive adenines in the loop region between helix $H_1$ and helix $H_3$ |
| $\Delta G(H1,H2)$ | Thermodynamic free-energy associated with helix $H_1$, helix $H_2$ and the loop region between $H_1$ and $H_2$ |
| $\Delta G(H2,H3)$ | Thermodynamic free-energy associated with helix $H_2$, helix $H_3$ and the loop region between $H_2$ and $H_3$ |
| $\Delta G(H1,H3)$ | Thermodynamic free-energy associated with helix $H_1$, helix $H_3$ and the loop region between $H_1$ and $H_3$ |

**Table 2.** Features Used for Predicting Coaxial Helical Stacking of Three-Way Junctions

| Feature Name | Feature Value |
|---|---|
| \|J12\| | 0 |
| \|J23\| | 4 |
| \|J13\| | 0 |
| Min(\|J12\|,\|J23\|,\|J13\|) | 0 |
| Med(\|J12\|,\|J23\|,\|J13\|) | 0 |
| Max(\|J12\|,\|J23\|,\|J13\|) | 4 |
| Min(\|J23\|,\|J13\|) | 0 |
| Min(\|J12\|,\|J13\|) | 0 |
| Min(\|J12\|,\|J23\|) | 0 |
| A(J12) | 0 |
| A(J23) | 1 |
| A(J13) | 0 |
| $\Delta G(H1,H2)$ | -1.4 |
| $\Delta G(H2,H3)$ | 6.3 |
| $\Delta G(H1,H3)$ | -2.1 |

**Table 3.** Fifteen Feature Values for the Three-Way Junction Illustrated in Figure 3

**The CSminer Approach**

We selected samples from known RNA junctions [31] with similar features. We obtained "true" secondary structure for each molecule from PDB [27] crystals (i.e. "the gold standard") using S2S [53]. We clustered these structures using RNAforester [58]. We manually constructed a Stockholm alignment (Figure 5). We created a covariance model from the Stockholm alignment using Infernal's CMbuild utility [19]. We modified the Infernal source code in the CMsearch Infernal utility to execute Junction Explorer [2] whenever a secondary structure similar to our covariance model was found in the genome.

Junction Explorer (JE) is an ensemble based classifier capable of predicting the type of coaxial helical stacking in an RNA junction based upon the secondary structure alone [2].

```
# STOCKHOLM 1.0
#=GF ID 1NKW(16) d. radiodurans length=114
#=GF ID 2AW4(18) e. coli length=114
#=GF ID 2J01(19) t. thermophilus length=114
#=GF alignment by S2S software

1NKW-16         GUGAGCUAAGCAUGACC-----AGGU----UGAAACCCCCGUGACAGGGGGCGGAG----
2AW4-18         GACCG-A-----ACC-GACUAAUGUUGAAAAAUUAGCGGAUGA-CUUGUGGCUGGGGGUG
2J01-19         GCGAGCUAGCCCUGGCCAGGGUGAAG----CU----GGGGUGAG--ACCC----AG----
#=GC SS_cons    ((.....(((((((((......(((....((.....(((......))))....))....

1NKW-16         GACCG-A---A-CCGGU-GCCUGCUGAAA--CAGUCUCGGAUGAGUUGUGUUUAGG-AGU
2AW4-18         GUGAUCUAGCCAUGGGC-----AGGUUGAAGG----UUGGGUAACACUAA----CUGGAG
2J01-19         -UGGAGGCCCGAACC-GGUGGGGGAUGCAAACCCCUCGGAUGA-GCUGGGGCUAGGAGUG
#=GC SS_cons    .))))......(((.(.(((((........)))))))))....))))))))))....((.

1NKW-16         GAAAAGCU-AACCGAAC
2AW4-18         AAAG--GCCAAUCAAAC
2J01-19         AAAA--GCUAACCGAGC
#=GC SS_cons    ......))......))
//
```

**Fig. (5).** Stockholm alignment of RNA structures from three organisms recorded in PDB with identifiers 1NKW, 2AW4 and 2J01.

JE is built using random forests and is comprised of numerous classification and regression trees (CARTs) [59], each of which is formed by a small random subset (i.e. the square root) of features determined to be the most important features describing the RNA junction. Each JE CART is capable of contributing a "better than random opinion" about the coaxial helical stacking classification of an unknown input. By consolidating all opinions from all CARTs, i.e. by tallying all "votes", JE is able to predict the coaxial helical stacking status of the RNA junction with greater accuracy than other existing comparable methods.

In the case of a 3-way junction, JE evaluates 15 features (see Table 2) readily available in the secondary structure information. Other higher order junctions require varying numbers of features to be evaluated. In this work, we only focused on the classification of 3-way junctions to simply demonstrate the viability of combining Infernal functionality with JE functionality in a genome wide search. The extension of the Infernal CMsearch utility with these modifications is a new program named CSminer.

**RESULTS**

We apply CSminer to H. marismortui chromosome II. Figure 6 illustrates the output of CSminer. The output produced by CSminer is restricted to only those results determined to contain a multi-branch loop. Furthermore, in each case of a multi-branch loop, a "Coaxial Stack Status" is reported.

Figure 6 shows that in the target genome, H. marismortui chromosome II, there is evidence of two 3-way junctions. Furthermore, each of these 3-way junctions, upon immediate analysis by Junction Explorer, is predicted to contain a coaxial helical stacking. One coaxial helical stacking is of type H1H2 (i.e. helix H1 and helix H2 are aligned with a common axis) and the second is of type H2H3 (i.e. helix H2 and helix H3 are aligned with a common axis). Figure 7 shows the primary sequence of the search result from the genome. The three components that comprise a 3-way junction, as described previously, are highlighted for clarification for each of the two 3-way junctions detected. One 3-way junction is highlighted in yellow and the second 3-way junction is highlighted in green. Figure 8 illustrates these two 3-way junctions in a 2D plot with each 3-way junction enclosed by a red dotted line. The yellow highlighted 3-way junction from Figure 7 is identified as "A" in Figure 8, and the green 3-way junction from Figure 7 is identified as "B" in Figure 8.

```
CM: 2J01(19)
>gi|55380074|ref|NC_006397.1|

 Minus strand results:

 Query = 1 - 107, Target = 2771 - 2656
 Score = 47.55, GC =  51

 Coax status = H1H2, H2H3

         [[,,,,,(((((((((,<<<<.....<<<<<<______>>>>>>.....->>>>,,<<<<
        1 GcGAgCUAgcCauGgcCaggu.....cgccggguAACAccggcg.....GaccgAacCcg 50
          :CGA CUA::CA:GG:CA:G:     CG::::G AA ::::CG     G:C:G   ::G
     2771 ACGAUCUACGCAUGGACAAGAugaagCGUGCCGAAA--GGCACGuggaaGUCUGUUAGAG 2714

          -<<<.<<________>>>>>>>>>,,,)))))))))),,,,<<______>>,,,,,,,]]
       51 aggg.ggguugAAAAcccccccgGgugAgcUguGgcUAGGagggAAAAccuAAcCgAgC 107
            GG G:   U+ AA :CCC C:: UGA:CU:UG::UAGG+G:GAAA  :C+ A+CGAG:
     2713 UUGGuGUCCUACAAUACCCUCUCGUGAUCUAUGUGUAGGGGUGAAAGGCCCAUCGAGU 2656
```

**Fig. (6).** CSminer's search result from genome H. marismortui chromosome II. Nucleotides shown are from positions 2771 through 2656, i.e. for a length of 116, on the minus strand. Notice the coaxial helical stacking status of "H1H2, H2H3" indicating that evidence of two 3-way junctions was located and that each 3-way junction is predicted to contain a coaxial helical stacking.

```
ACGAUCUACGCAUGGACAAGAUGAAGCGUGCCGAAAGGCACGUGGAAGUCUGUUAGAGUUGGUGUCCUACA
            AUACCCUCUCGUGAUCUAUGUGUAGGGGUGAAAGGCCCAUCGAGU
```

**Fig. (7).** Primary sequence of nucleotides 2771 through 2656 on the minus strand of genome H. marismortui chromosome II. Highlighted in yellow are the subsequences that comprise the 3-way junction "A" in Figure 8. Highlighted in green are the subsequences that comprise the 3-way junction "B" in Figure 8.

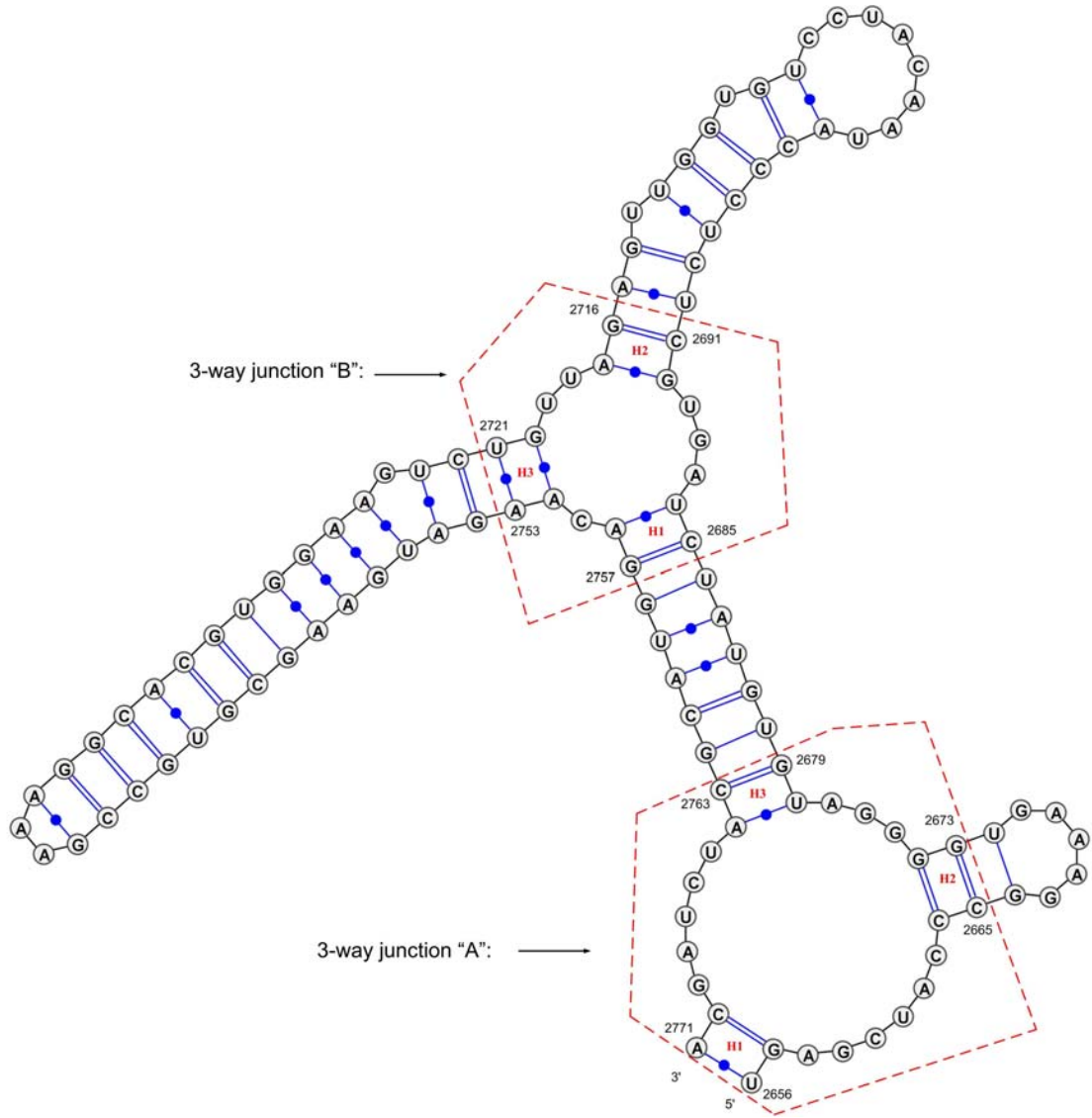

**Fig. (8).** 2D plot produced by VARNA [54] of nucleotides 2771 through 2656 (i.e. minus strand) of genome H. marismortui chromosome II using CT format provided by CSminer (see Figure 6). Two 3-way junctions are enclosed by dotted red lines.

This CSminer search result is confirmed as follows. We downloaded the chain 0 nucleotide FASTA sequence from PDB [27] for the 1S72 structure. Using NCBI Blast [60], we located this downloaded FASTA sequence in chromosome II of H. marismortui, i.e. GenBank ID AY596298.1 from position 3,542 through 2,922 on the negative strand. Therefore, we confirmed that the RNA crystal structure is located in genome H. marismortui chromosome II. The CSminer search was performed on this genome and a match was found between positions 2771 and 2656 on the minus strand (see Figure 6), i.e. within coordinates of the Blast search result above.

To further verify the results from CSminer, the Junction Explorer webserver available at http://bioinformatics.njit.edu/junction/ was run using the secondary structure produced by CSminer in CT format. As expected, two 3-way junctions were predicted by the Junction Explorer webserver, i.e. types H1H2 and H2H3. These two results are illustrated in Figure 9. Notice that the results produced by the Junction Explorer webserver do not reflect the correct genomic coordinates from H. marismortui chromosome II because this information was not provided within the CT formatted file. Nevertheless, all nucleotide content and structure information is preserved and correctly illustrated in the two 3-way junctions represented in Figure 9.

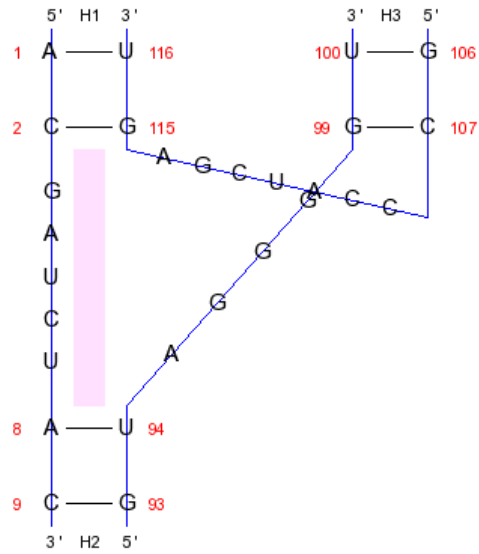

(a)

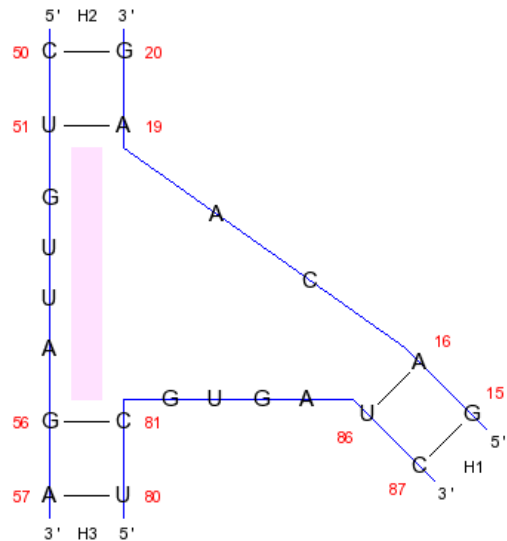

(b)

**Fig. (9).** Graphical results from the Junction Explorer webserver [2] which predicts an H1H2 coaxial helical stacking and an H2H3 coaxial helical stacking in the RNA sequence located by CSminer (Figure 6). (a) represents the 3-way junction "A" in Figure 8. (b) represents the 3-way junction "B" in Figure 8.

**DISCUSSION**

CSminer combines the strengths of the genome search tool, Infernal, with an innovative ensemble based classifier, Junction Explorer, to provide an effective new predictive instrument. Among the growing number of ensemble based methodologies, the random forests method, utilized by Junction Explorer, is among the most accurate. This allows us to add significant additional functionality to Infernal. Efficient prediction of the presence of coaxial helical stacking motifs in genomes will help to further unravel the mysteries of noncoding RNA. Much remains unknown in this exciting research area. Progress is measured in small, but positive, achievements, such as our modest contribution described in this work. Our conclusion is that genome-wide searching for coaxial helical stacking RNA motifs is feasible and cost effective.

**CURRENT & FUTURE DEVELOPMENTS**

We are currently conducting more extensive CSminer genome searches by using known higher order junctions (i.e. 4-way junctions, 5-way junctions, etc.) to further demonstrate the feasibility of this novel approach. We are also investigating the use of various ensemble based and feature selection methodologies, including random forests, to find RNA tertiary motifs which include pseudoknot interactions and A-minors [12]. We are interested in evaluating the effectiveness and efficiency of combining a variety of different machine learning approaches to stubborn RNA motif finding problems.